\documentclass[runningheads,citeauthoryear]{apinv}
\usepackage{epsfig,cite,graphics}
\usepackage[T2A]{fontenc}
\usepackage[cp1251]{inputenc}

\begin{document}

\title{Long-term $BVRI$ light curves of 5 pre-main sequence stars in the field of "Gulf of Mexico"}
\titlerunning{Long-term $BVRI$ light curves of 5 PMS stars in "Gulf of Mexico"}
\author{Sunay I. Ibryamov\inst{1}, Evgeni H. Semkov\inst{1}, Stoyanka P. Peneva\inst{1}}
\authorrunning{S. Ibryamov et al.}
\tocauthor{S. Ibryamov et al.} 
\institute{Institute of Astronomy and National Astronomical Observatory, Bulgarian Academy of Sciences, 72, Tsarigradsko Shose Blvd., 1784 Sofia, Bulgaria
	\newline
	\email{sibryamov@astro.bas.bg}}
\papertype{Submitted on; Accepted on}	
\maketitle

\begin{abstract}
We present new data from $BVRI$ photometric observations of five PMS stars during the period from April 2013 to July 2014. The stars are located in the field of NGC 7000/IC 5070 ("Gulf of Mexico") -- a region with active star formation. The presented paper is a continuation of our long-term photometric investigations of the young stellar objects in this region. The long-term multicolor photometric observations of PMS stars are very important for their exact classification. Our results show that the studied stars exhibit different types of photometric variability in all bands. We tried to classify them using our data from the long-term photometry and data published by other authors.
\end{abstract}

\keywords{stars: pre-main sequence, stars: variables: T Tauri, UX Orionis, star: individual: V521 Cyg, V752 Cyg, V1539 Cyg, V1716 Cyg, V2051 Cyg}

\section*{1. Introduction}

The Pre-Main Sequence (PMS) stars are among the most studied objects in the modern astrophysics. The studies of PMS stars give us information for the early stages of stellar evolution and an opportunity to test stellar evolution scenarios. 

Photometric variability in PMS stars is a common phenomenon. Both classes of PMS stars -- low-mass (M$\leq$2M$_{\sun}$) T Tauri stars (TTS) and the more massive (2M$_{\sun}$<M<8M$_{\sun}$) Herbig Ae/Be stars (HAEBES) show different types of photometric variability (Herbst et al. 1994, 2007).

The study of TTS began after the work of Joy (1945). Their main characteristics are the irregular photometric variability and the emission line spectra. T Tauri stars are separated in two main subclasses: Classical T Tauri stars (CTTS) surrounded by spacious circumstellar disks and Weak-line (also called "naked") T Tauri stars (WTTS) without evidence for disk (Bertout 1989).

Herbst et al. (2007) defined five types of brightness variation concerning PMS stars. The variability of Type I is due to cool spots or groups of spots on the stellar surface. This light variability with amplitudes about 0.03-0.3 mag, and in extreme cases reaching 0.8 mag in the $V$-band is typical for WTTS, but rarely that can also be observed in the CTTS. The Type II of variability shows often irregular variations with larger photometric amplitudes (2-3 mag), associated with highly variable accretion from the circumstellar disk onto the surface of CTTS. The variability of Type III is due to rotating hot spots on stellar surface and only seen in CTTS. Periodicity in this type of variability can be observed only for a couple of rotation cycles. The variability of Type IV is due to flare-like variations, which are typical for WTTS in $B$ and $U$-band. Flares are random with different sized amplitudes, as there was no periodicity. Type V of variability is characterized with brightness dips lasting from few days up to several months, which presumably result from circumstellar dust or clouds obscuration. This type of variability is commonly observed in early type TTS and HAEBES and shows large photometric amplitudes (up to 2.8 mag in $V$-band). The prototype of this group of PMS objects with intermediate mass named UXors is UX Orionis (Grinin et al. 1991).

The large amplitude outburst of PMS stars are grouped into two main types named after their prototypes: FU Orionis (FUor; Ambartsumian 1971) and EX Lupi (EXor; Herbig 1989). During the quiescence state FUor and EXor stars are probably normally accreting TTS with massive circumstellar disks. The outbursts of both types of variables are generally attributed to a sizable increase in accretion rate from the circumstellar disk onto the stellar surface. The outbursts of FUor objects with amplitude over 5 mag are very rare, and the rise of brightness is shorter than the decline, while EXor objects show frequent (every few years or a decade) irregular or relatively brief (a few months to one year) outburst with an amplitude of several magnitudes (up to 5 mag).

The five stars from our study are located in the field of "Gulf of Mexico", near to the new FUor star V2493 Cyg erupted in 2010 (Semkov et al. 2010, 2012, Miller et al. 2011). "Gulf of Mexico" is a region with active star formation (Armond et al. 2011) located between the North America Nebula (NGC 7000) and Pelican Nebula (IC 5070). These nebulae are part of a single large HII region W80. The "Gulf of Mexico" is rich in young stellar objects, as H$\alpha$ emission line stars, flare stars from UV Ceti type, TTS and HAEBES. Recently, the results of two extensive photometric studies of PMS stars in this field (Findeisen et al. 2013, Poljan\v{c}i\'{c} Beljan et al. 2014) have been published. 

\section*{2. Observations and data reduction}

The photometric $BVRI$ data presented in this paper were collected in the period from April 2013 to July 2014. The CCD observations were carried out in two observatories with four telescopes: the 2-m Ritchey-Chr\'{e}tien-Coud\'{e} (RCC), the 50/70-cm Schmidt and the 60-cm Cassegrain telescopes of the Rozhen National Astronomical Observatory (Bulgaria) and the 1.3-m Ritchey-Chr\'{e}tien (RC) telescope of the Skinakas Observatory\footnote{Skinakas Observatory is a collaborative project of the University of Crete, the Foundation for Research and Technology, Greece, and the Max-Planck-Institut f{\"u}r Extraterrestrische Physik, Germany.} of the University of Crete (Greece).

The observations were performed with four types of CCD cameras: VersArray 1300B at the 2-m RCC telescope, ANDOR DZ436-BV at the 1.3-m RC telescope, FLI PL16803 at the 50/70-cm Schmidt telescope, and FLI PL09000 at the 60-cm Cassegrain telescope. The technical parameters and specifications for the cameras used are summarized in Table 1.

\begin{table}[htb!!!]
  \begin{center}
  \caption{CCD cameras technical parameters and specifications}
  \begin{tabular}{llccccc}
	  \hline\hline
	  \noalign{\smallskip}
  Telescope & CCD Camera & Chip size [\textit{pix}] & Pixel size & RON & Gain & Scale\\
    \noalign{\smallskip}
	  \hline
	  \noalign{\smallskip}
  2-m RCC & VersArray 1300B & 1340 $\times$ 1300 & 20 $\mu m$ & 2.0 e$^{-}$ rms & 1.0 e$^{-}$/ADU & 0.26\arcsec/pix \\
	1.3-m RC & ANDOR DZ436-BV & 2048 $\times$ 2048 & 13.5 $\mu m$ & 8.14 e$^{-}$ rms & 2.7 e$^{-}$/ADU & 0.28\arcsec/pix\\
	Schmidt & FLI PL16803 & 4096 $\times$ 4096 & 9 $\mu m$ & 9.0 e$^{-}$ rms & 1.0 e$^{-}$/ADU & 1.08\arcsec/pix\\
	Cassegrain & FLI PL09000 & 3056 $\times$ 3056 & 12 $\mu m$ & 8.5 e$^{-}$ rms & 1.0 e$^{-}$/ADU & 0.33\arcsec/pix\\
	  \hline
  \end{tabular}
  \label{table1}
  \end{center}
\end{table} 

All frames were taken through a standard Johnson-Cousins set of filters. Twilight flat-fields in each filter were obtained each clear evening or morning. All frames obtained with the VersArray 1300B and ANDOR DZ436-BV cameras are bias subtracted and flat-field corrected. CCD frames obtained with the FLI PL16803 and FLI PL09000 cameras are dark-frame subtracted and flat-field corrected. The standard IDL procedures (adapted from $DAOPHOT$) were used for reduction of the photometric data. All data were analyzed using the same aperture, which was chosen to have a 4$\arcsec$ radius, while the background annulus was taken from 9$\arcsec$ to 14$\arcsec$. As a reference, the $BVRI$ comparison sequence reported in Semkov et al. (2010) was used. The average value of the errors in the reported magnitudes are 0.01-0.02 mag for $I$ and $R$-band data, 0.02-0.05 mag for $V$-band data, and 0.02-0.09 mag for $B$-band data.

\section*{3. Results}

In the present paper we report recent photometric data from $BVRI$ CCD observations of five PMS stars: V521 Cyg, V752 Cyg, V1539 Cyg, V1716 Cyg, and V2051 Cyg. 
Our data are a continuation of the long-term photometric investigations of these stars published in Poljan\v{c}i\'{c} Beljan et al. (2014). Figure 1 shows color image of a part from the region of "Gulf of Mexico" where the stars from our study and the FUor star V2493 Cyg are marked. 

\begin{figure}[htb!!!]
  \begin{center}
    \centering{\epsfig{file=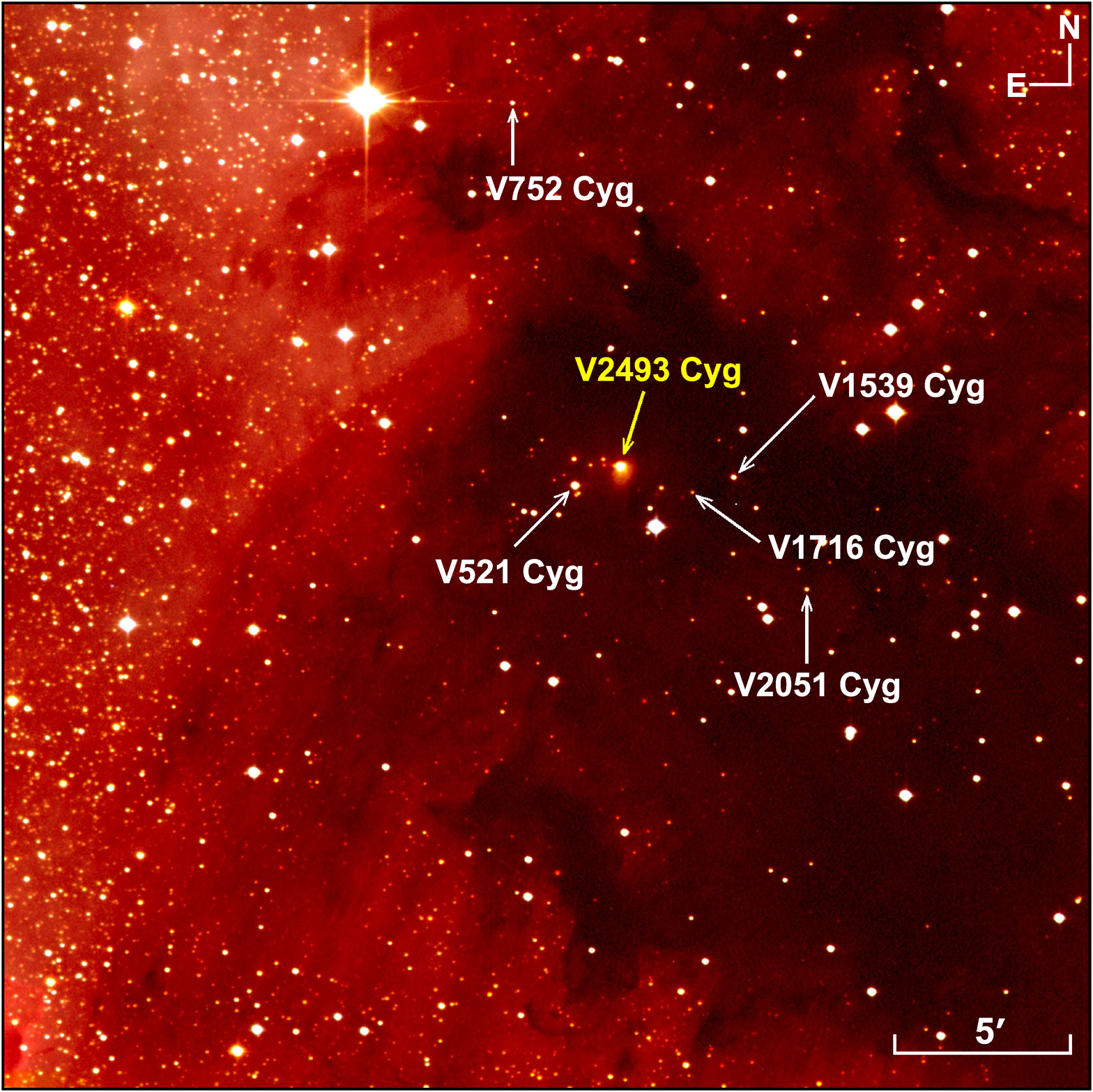, width=0.55\textwidth}}
    \caption[]{Color image of the region "Gulf of Mexico" obtained with 50/70-cm Schmidt telescope of Rozhen NAO. The stars from our study and the FUor star V2493 Cyg are marked on the image.}
    \label{countryshape}
  \end{center}
\end{figure}

\subsection*{3.1. V521 Cyg} 

V521 Cyg was discovered by Herbig (1958) as H$\alpha$ emission line star and classified as TTS by Herbig \& Bell (1988), Fernandez et al. (1995) and Laugalys et al. (2006). Terranegra et al. (1994) and Armond et al. (2011) classified the star as CTTS. Grankin et al. (2007) showed long-term photometric curve of V521 Cyg and concluded that the star exhibit unusual colour behavior with a blue turnaround at minimum brightness, probably caused from scattered light during partial occultation of the stellar photosphere by circumstellar material. Poljan\v{c}i\'{c} Beljan et al. (2014) determined the period of V521 Cyg, which is found to be 503 days.

The $BVRI$ light curves of V521 Cyg from all our CCD observations (Poljan\v{c}i\'{c} Beljan et al. 2014, and the present paper) are shown in Fig. 2. On the figure, circles denote photometric data acquired with the 2-m RCC telescope; diamonds - the photometric data taken with the 1.3-m RC telescope; triangles - the photometric data collected with the 50/70-cm Schmidt telescope, and squares - the photometric data obtained with the 60-cm Cassegrain telescope. 

The photometric results of our recent CCD $BVRI$ observations of V521 Cyg are summarized in Table 2. The columns contains Julian Date (J.D.) of observations, measured $IRVB$ magnitudes of the star, telescope and CCD camera used. 

\begin{figure}[htb!!!]
  \begin{center}
    \centering{\epsfig{file=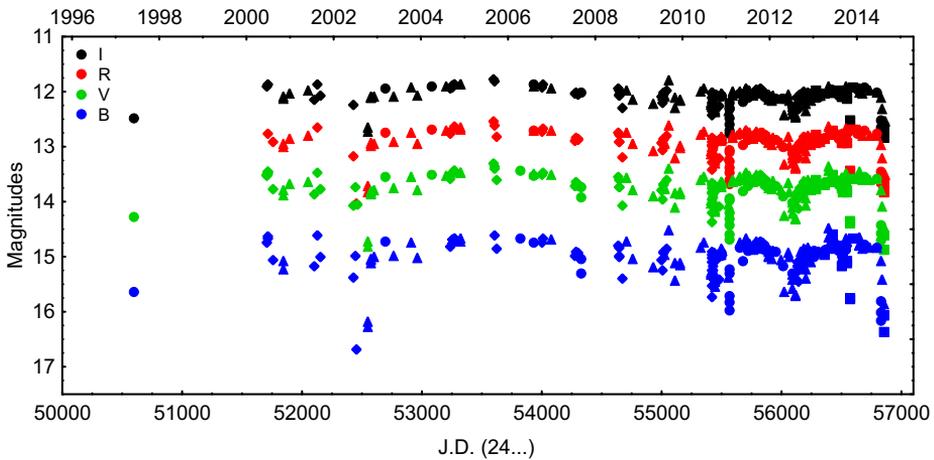, width=0.95\textwidth}}
    \caption[]{$BVRI$ light curves of V521 Cyg for the period June 1997 -- July 2014}
    \label{countryshape}
  \end{center}
\end{figure}

\begin{table}[htb!!!]
  \begin{center}
  \caption{Photometric CCD observations of V521 Cyg during the period April 2013 -- July 2014}
  \begin{tabular}{cccccccccccccc}
		  \hline \hline
		  \noalign{\smallskip}
J.D. (24...) & $I$ & $R$ & $V$ & $B$ & Tel & CCD & J.D. (24...) & $I$ & $R$ & $V$ & $B$ & Tel & CCD\\
  \noalign{\smallskip}
  \hline 
  \noalign{\smallskip}
56392.510 &	12.02 &	12.74 &	13.45 &	14.66 &	Sch &	FLI & 56544.264	& 12.09	& 12.89	& 13.69	& 14.89	& 2-m	& VA\\
56394.482	& 11.92 &	12.64 &	13.37	& 14.49 &	Sch &	FLI & 56547.383	& 12.12	& 12.94	& 13.81	& 15.09	& Cas	& FLI\\
56415.418	& 11.99 &	12.79 &	13.61 &	14.73	& Sch &	FLI & 56550.363	& 12.14	& 12.94	& 13.84	& 15.13	& Cas	& FLI\\
56428.404 &	12.00 &	12.78	& 13.62 &	14.76	& Cas	& FLI & 56553.283	& 12.00	& 12.79	&	- & 14.80	& 1.3-m	& AND\\
56430.408	& 11.98	& 12.72	& 13.52	& 14.61	& Cas	& FLI & 56577.306	& 12.55	& 13.49	& 14.40	& 15.79	& Cas	& FLI\\
56432.405	& 12.02	& 12.75	& 13.66	& 14.86	& Cas	& FLI & 56578.331	& 12.53	& 13.44	& 14.36	& 15.76	& Cas	& FLI\\
56443.367	& 12.02	& 12.77	& 13.55	& 14.74	& Cas	& FLI & 56604.264	& 11.94	& 12.69	& 13.60	& 14.68	& Cas	& FLI\\
56444.354	& 12.02	& 12.77	& 13.58	& 14.68	& Cas	& FLI & 56636.185	& 11.97	& 12.77	& 13.57	&	- & 2-m & VA\\
56478.363	& 12.09	& 12.87	& 13.70	& 14.91	& 2-m	& VA  & 56655.200	& 11.96	& 12.70	& 13.52	& 14.79	& Sch	& FLI\\
56507.273	& 12.04	& 12.92	& 13.74	&	- & 2-m & VA      & 56656.180	& 11.95	& 12.72	& 13.47	& 14.66	& Sch	& FLI\\
56508.318	& 12.10	& 12.91	& 13.72	& 14.94 &	2-m	& VA  & 56657.193	& 12.02	& 12.74	& 13.55	& 14.80	& Sch	& FLI\\
56509.288	& 11.99	& 12.78	& 13.58	& 14.83	& Sch	& FLI & 56681.189	& 12.05	& 12.82	& 13.62	& 14.87 &	Sch	& FLI\\
56510.369	& 12.01	& 12.82	& 13.64	& 14.82	& Cas	& FLI & 56694.194	& 11.94	& 12.72	& 13.55	& 14.79	& 2-m	& VA\\
56510.386	& 12.01	& 12.78	& 13.57	& 14.83	& Sch	& FLI &	56738.590	& 11.98	& 12.81	& 13.62	& 14.85	& Sch	& FLI\\
56511.411	& 12.10	& 12.91	& 13.72	& 14.94	& Sch	& FLI & 56799.425	& 12.02	& 12.81	& 13.60 & 14.84	& Sch	& FLI\\
56511.413	& 12.11	& 12.95	& 13.77	& 15.18	& Cas	& FLI & 56801.436	& 12.03	& 12.78	& 13.59	& 14.84	& 2-m	& VA\\
56512.398	& 12.02	& 12.83	& 13.63	& 14.88	& Sch	& FLI & 56832.391	& 12.73	& 13.68	& 14.71	& 16.18	& 2-m	& VA\\
56512.403	& 12.01	& 12.82	& 13.75	& 14.81	& Cas	& FLI & 56834.356	& 12.66	& 13.59	& 14.58	& 16.03	& 2-m	& VA\\
56513.382	& 11.98	& 12.79	& 13.64	& 14.89	& Cas	& FLI & 56835.462	& 12.54	& 13.47	& 14.43	& 15.81	& 2-m	& VA\\
56514.350	& 12.09	& 12.95	& 13.83	& 15.01	& Cas	& FLI & 56837.417	& 12.13	& 12.98	& 13.81	& 15.09	& Sch	& FLI\\
56517.284	& 12.02	& 12.80	& 13.63	& 14.87	& Cas	& FLI & 56838.398	& 12.32	& 13.23	& 14.10	& 15.43	& Sch	& FLI\\
56540.274	& 11.93	& 12.68	& 13.50	& 14.67	& Sch	& FLI & 56859.390 & 12.85 & 13.84 & 14.89 & 16.38 & Cas & FLI\\
56541.322	& 11.99	& 12.78	& 13.60	& 14.83	& Sch	& FLI & 56860.391 & 12.67 & 13.67 & 14.63 & 16.08 & Cas & FLI\\
56542.381	& 11.92	& 12.70	& 13.51	& 14.72	& Sch	& FLI & 56863.339	& 12.56	& 13.53	& 14.47	& 15.88	& Sch	& FLI\\
56543.396	& 12.01	& 12.79	& 13.59	& 14.80	& 2-m	& VA  & 56864.356	& 12.55	& 13.52 & - & - &	Sch &	FLI\\
\hline
  \end{tabular}
  \label{table2}
  \end{center}
\end{table} 

The brightness variations of V521 Cyg in the different bands during the period of our study (1997-2014) are 11.78 -- 12.85 mag for $I$-band, 12.55 -- 14.05 mag for $R$-band, 13.32 -- 14.89 mag for $V$-band, and 14.49 -- 16.69 mag for $B$-band. The observed amplitudes are 1.07 mag for $I$-band, 1.50 mag for $R$-band, 1.57 mag for $V$-band and 2.20 mag for $B$-band in the same period.

Figure 2 shows the long-term photometric variability of V521 Cyg and the presence of seven deep declines in brightness observed in all bands, as follows: two deep declines observed in 2002, one in 2010, one decline in the beginning of 2011, one lasting longer decline in 2012, one at the end of 2013, and one very deep decline in 2014. During these declines, the brightness of the star faded for more that 1 mag. There may be others deep declines of brightness, but they were not registered in our photometric study. The declines in brightness of V521 Cyg likely are caused by partial and irregular obscuration of the star by circumstellar material and V521 Cyg can be classified as CTTS with evidences for Type V (UXor-type) of variability. The results obtained in our study support the results of Grankin et al. (2007) about the causes of variability of V521 Cyg. But the different shape of the observed declines gives grounds to predict a variety of eclipse reasons.

\subsection*{3.2. V752 Cyg}

The variability of V752 Cyg was discovered by Erastova \& Tsvetkov (1978) and confirmed by Kohoutek \& Wehmeyer (1997). For the first time the long-term $BVRI$ photometric light curves of V752 Cyg were published in Poljan\v{c}i\'{c} Beljan et al. (2014).

The recent photometric results from our CCD $BVRI$ observations of the star are summarized in Table 3. The columns have the same contents as in Table 2. The $BVRI$ light curves of V752 Cyg during the period of all our observations (Poljan\v{c}i\'{c} Beljan et al. 2014, and the present paper) are plotted in Fig. 3. The symbols used for different telescopes are as in Fig. 2. The figure shows very strong and irregular variability, without evidence of periodicity. The variations in brightness of V752 Cyg in the different bands durind the period of our photometric study (2006-2014) are 14.28 -- 15.55 mag for $I$-band, 14.89 -- 16.67 mag for $R$-band, 15.38 -- 17.71 mag for $V$-band, and 16.02 -- 18.89 mag for $B$-band. The observed amplitudes are 1.27 mag for $I$-band, 1.78 mag for $R$-band, 2.33 mag for $V$-band and 2.87 mag for $B$-band in the same period.

\begin{figure}[htb!!!]
  \begin{center}
    \centering{\epsfig{file=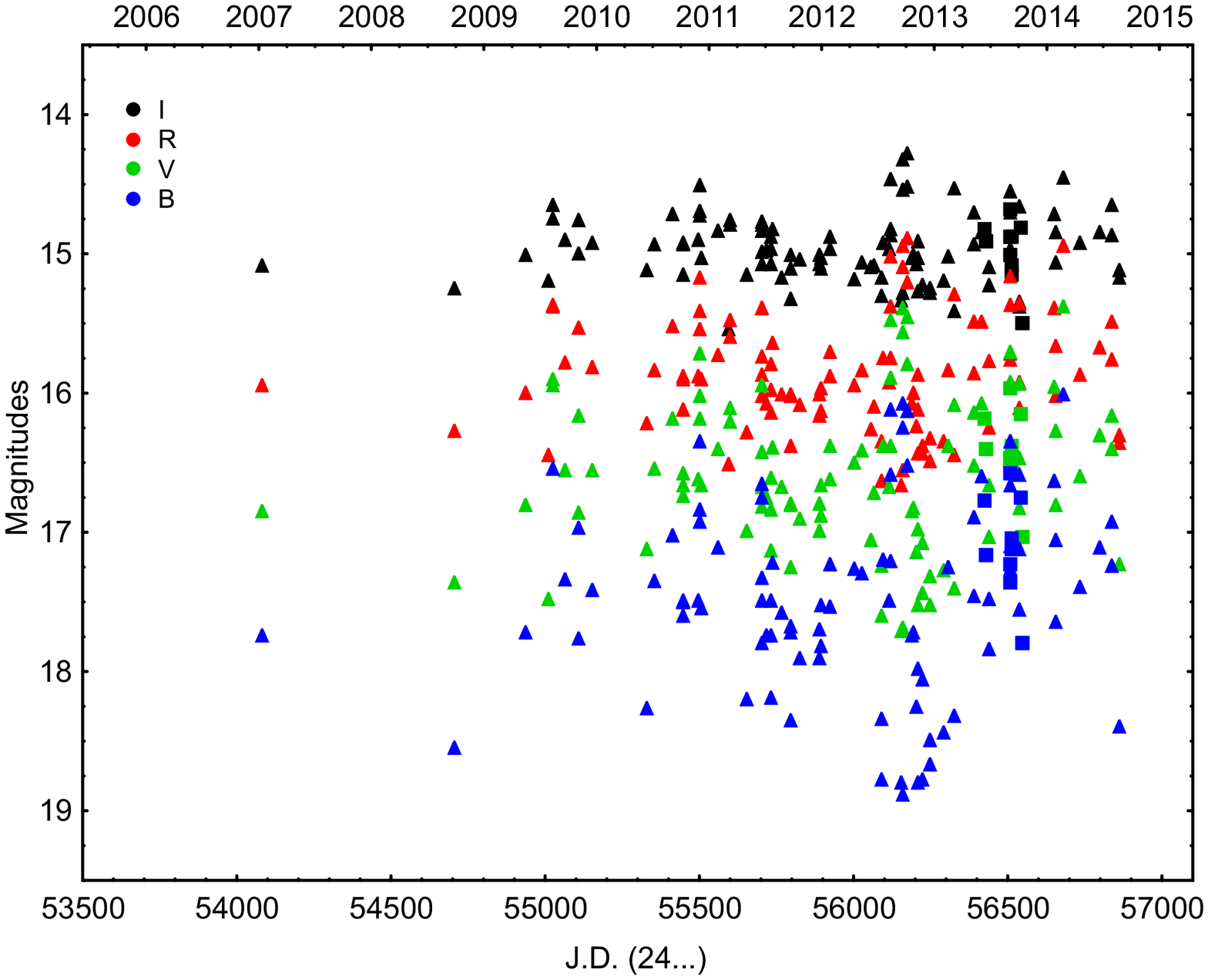, width=0.9\textwidth}}
    \caption[]{$BVRI$ light curves of V752 Cyg for the period December 2006 -- July 2014}
    \label{countryshape}
  \end{center}
\end{figure}

\begin{table}[htb!!!]
  \begin{center}
  \caption{Photometric CCD observations of V752 Cyg during the period April 2013 -- July 2014}
  \begin{tabular}{cccccccccccccc}
		  \hline \hline
		  \noalign{\smallskip}
J.D. (24...) & $I$ & $R$ & $V$ & $B$ & Tel & CCD & J.D. (24...) & $I$ & $R$ & $V$ & $B$ & Tel & CCD\\
  \noalign{\smallskip}
  \hline 
  \noalign{\smallskip}
56392.510 & 14.94	& 15.86	& 16.53	& 17.46	& Sch	& FLI & 56517.284	& 15.09	& 15.86	& 16.53	& 17.05	& Cas & FLI\\
56394.482	& 14.71	& 15.49	& 16.15	& 16.90	& Sch	& FLI & 56540.274	& 14.66	& 15.36	& 15.94	& 16.59	& Sch	& FLI\\
56415.418	& 14.85	& 15.49	& 16.08	& 16.60 &	Sch	& FLI & 56541.322	& 15.35	& 15.93	& 16.47	& 17.13	& Sch	& FLI\\
56428.404	& 14.83	& 15.58	& 16.19	& 16.78	& Cas & FLI & 56542.381	& 15.38	& 16.11	& 16.83	& 17.56	& Sch	& FLI\\
56430.408	& 14.91	& 15.71	& 16.41	& 17.17	& Cas & FLI & 56547.383	& 14.82	& 15.51	& 16.16	& 16.76	& Cas	& FLI\\
56443.367	& 15.10 & 15.78	& 16.67	& 17.48	& Sch	& FLI & 56550.363	& 15.50 & 16.20 & 17.04	& 17.80	& Cas	& FLI\\
56444.354	& 15.23	& 16.26	& 17.04	& 17.85	& Sch	& FLI & 56655.200 & 14.72	& 15.40	& 15.96	& 16.64	& Sch	& FLI\\
56509.288	& 14.56	& 15.17	& 15.71	& 16.35	& Sch	& FLI & 56656.180	& 14.85	& 15.67	& 16.28	& 17.06	& Sch	& FLI\\
56510.369	& 14.69	& 15.40 & 15.97 & 16.58	& Cas & FLI & 56657.193	& 15.07	& 16.03	& 16.81	& 17.65	& Sch	& FLI\\
56510.386	& 14.71	& 15.37	& 15.93	& 16.67	& Sch	& FLI & 56681.189	& 14.46	& 14.95	& 15.38	& 16.02	& Sch	& FLI\\
56511.411	& 14.97	& 15.72	& 16.35	& 17.10 & Sch	& FLI & 56738.590	& 14.93	& 15.87	& 16.60	& 17.40	& Sch	& FLI\\
56511.413	& 15.01	& 15.81	& 16.47	& 17.23	& Cas & FLI & 56799.425	& 14.85	& 15.68	& 16.31	& 17.11	& Sch	& FLI\\
56512.398	& 14.88	& 15.76	& 16.47	& 17.27	& Sch	& FLI & 56837.417	& 14.87	& 15.76	& 16.41	& 17.25	& Sch	& FLI\\
56512.403	& 14.88	& 15.74	& 16.49	& 17.36	& Cas & FLI & 56838.398	& 14.65	& 15.49	& 16.17	& 16.93	& Sch	& FLI\\
56513.382	& 14.88	& 15.69	& 16.38	& 17.12	& Cas & FLI & 56863.339	& 15.18	& 16.31	& 17.23	& 18.40 &	Sch	& FLI\\
56514.350 & 15.16	& 15.91	& 16.55	& 17.13	& Cas & FLI & 56864.356	& 15.12 &	16.36 & - & - &	Sch	& FLI\\
\hline
  \end{tabular}
  \label{table3}
  \end{center}
\end{table} 

The irregular variations in brightness of V752 Cyg and the large photometric amplitudes (up to 2.33 in $V$-band) are indication, that the star can be classified as CTTS with Type II of variability. The observed strong photometric variability could be caused by the highly variable accretion from the circumstellar disk onto the stellar surface.

\subsection*{3.3. V1539 Cyg}

V1539 Cyg was discovered by Herbig (1958) as H$\alpha$ emission line star and classified as TTS by Herbig \& Bell (1988). Armond et al. (2011) classified V1539 Cyg as CTTS.

The $BVRI$ light curves of V1539 Cyg from all our CCD observations (Poljan\v{c}i\'{c} Beljan et al. 2014, and the present paper) are shown in Fig. 4.
The symbols used for different telescopes are as in Fig. 2. Table 4 presented our recent CCD $BVRI$ observations of the star. The columns have the same contents as in Table 2.

\begin{figure}[htb!!!]
  \begin{center}
    \centering{\epsfig{file=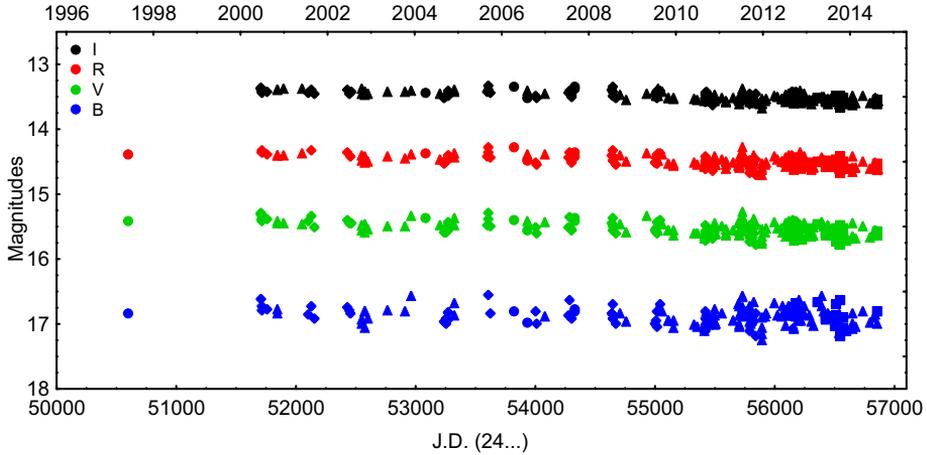, width=0.95\textwidth}}
    \caption[]{$BVRI$ light curves of V1539 Cyg for the period June 1997 -- July 2014}
    \label{countryshape}
  \end{center}
\end{figure}

\begin{table}[htb!!!]
  \begin{center}
  \caption{Photometric CCD observations of V1539 Cyg during the period April 2013 -- July 2014}
  \begin{tabular}{cccccccccccccc}
		\hline \hline
		\noalign{\smallskip}
J.D. (24...) & $I$ & $R$ & $V$ & $B$ & Tel & CCD & J.D. (24...) & $I$ & $R$ & $V$ & $B$ & Tel & CCD\\
  \noalign{\smallskip}
  \hline 
  \noalign{\smallskip}
56392.510	& 13.52	& 14.44	& 15.44	& 16.58	& Sch	& FLI & 56542.381	& 13.62	& 14.62	& 15.77	& 17.14	& Sch	& FLI\\
56394.482	& 13.51	& 14.47	& 15.46	& 16.73	& Sch	& FLI & 56547.383	& 13.67	& 14.68	& 15.78	& 17.19	& Cas	& FLI\\
56415.418	& 13.51	& 14.49	& 15.50	& 16.82	& Sch	& FLI & 56550.363	& 13.5	& 14.43	& 15.49	& 16.64	& Cas	& FLI\\
56428.404	& 13.59	& 14.58	& 15.65	& 16.94	& Cas	& FLI & 56553.283	& 13.57	& 14.61	& - & 17.04	& 1.3-m	& AND\\
56430.408	& 13.58	& 14.55	& 15.60	& 16.92	& Cas	& FLI & 56577.306	& 13.62	& 14.62	& 15.67	& 17.06	& Cas	& FLI\\
56432.405	& 13.56	& 14.49	& 15.60	& 16.90	& Cas	& FLI & 56578.331	& 13.53	& 14.50	& 15.54	& 16.91	& Cas	& FLI\\
56443.367	& 13.54	& 14.52	& 15.49	& 16.90	& Sch	& FLI & 56604.264	& 13.61	& 14.61	& 15.72	& 17.12	& Cas	& FLI\\
56444.354	& 13.62	& 14.59	& 15.65	& 16.85	& Sch	& FLI & 56655.200	& 13.61	& 14.60	& 15.69	& 17.01	& Sch	& FLI\\
56509.288	& 13.51	& 14.51	& 15.54	& 16.92	& Sch	& FLI & 56656.180	& 13.64	& 14.66	& 15.66	& 17.09	& Sch	& FLI\\
56510.369	& 13.55	& 14.59	& 15.62	& 17.04	& Cas	& FLI & 56657.193	& 13.50	& 14.43	& 15.46	& 16.85	& Sch	& FLI\\
56510.386	& 13.55	& 14.53	& 15.57	& 16.96	& Sch	& FLI & 56681.189	& 13.61	& 14.60	& 15.69	& 17.05	& Sch	& FLI\\
56511.411	& 13.53	& 14.52	& 15.55	& 16.87	& Sch	& FLI & 56738.590	& 13.50	& 14.51	& 15.51	& 16.81	& Sch	& FLI\\
56511.413	& 13.56	& 14.57	& 15.59	& 16.93	& Cas	& FLI & 56799.425	& 13.60	& 14.61	& 15.68	& 16.99	& Sch	& FLI\\
56512.398	& 13.58	& 14.58	& 15.65	& 16.99	& Sch	& FLI & 56837.417	& 13.54	& 14.54	& 15.58	& 16.91	& Sch	& FLI\\
56512.403	& 13.59	& 14.57	& 15.74	& 16.87	& Cas	& FLI & 56838.398	& 13.53	& 14.53	& 15.57	& 16.86	& Sch	& FLI\\
56513.382	& 13.55	& 14.55	& 15.58	& 17.01	& Cas	& FLI & 56859.390 & 13.56 & 14.54 & 15.64 & 16.82 & Cas & FLI\\
56514.350	& 13.49	& 14.48	& 15.54	& 16.70	& Cas	& FLI & 56860.391 & 13.62 & 14.63 & 15.58 & - & Cas & FLI\\
56517.284	& 13.57	& 14.53	& 15.61	& 16.91	& Cas	& FLI & 56863.339	& 13.54	& 14.55	& 15.58	& 17.00 &	Sch &	FLI\\
56540.274	& 13.51	& 14.47	& 15.54	& 16.86	& Sch	& FLI & 56864.356	& 13.57	& 14.62 & - & - &	Sch &	FLI\\
56541.322	& 13.55	& 14.54	& 15.59	& 16.99	& Sch	& FLI & & & & & & & \\
\hline
  \end{tabular}
  \label{table4}
  \end{center}
\end{table} 

The variations in brightness of V1539 Cyg in the different bands during the period of our photometric study (1997-2014) are 13.34 -- 13.68 mag for $I$-band, 14.28 -- 14.72 mag for $R$-band, 15.28 -- 15.79 mag for $V$-band, and 16.56 -- 17.26 mag for $B$-band. The observed amplitudes are 0.34 mag for $I$-band, 0.44 mag for $R$-band, 0.51 mag for $V$-band and 0.70 mag for $B$-band in the same period. The irregular photometric variability with relatively small amplitudes likely are caused by rotating cool and hot spots on the stellar surface. Therefore, the variability of V1539 Cyg can be attributed to Type I and III.
 
\subsection*{3.4. V1716 Cyg}

The variability of V1716 Cyg was discovered by Erastova \& Tsvetkov (1978). Findeisen et al. (2013) described two burst of the star, which are separated by 35 days. The first burst lasting 5-20 days, and second burst lasting 3 days. Poljan\v{c}i\'{c} Beljan et al. (2014) determined the period of V1716 Cyg, which is found to be 4.15 days.
The periodicity is stable for a period of several years and is probably connected with the stellar rotation.

The $BVRI$ light curves of V1716 Cyg from all our CCD observations (Poljan\v{c}i\'{c} Beljan et al. 2014, and the present paper) are shown in Fig. 5.
The symbols used for different telescopes are as in Fig. 2. The photometric results of our CCD $BVRI$ observations of the star are summarized in Table 5. The columns have the same contents as in Table 2.

\begin{figure}[!!!htb]
  \begin{center}
    \centering{\epsfig{file=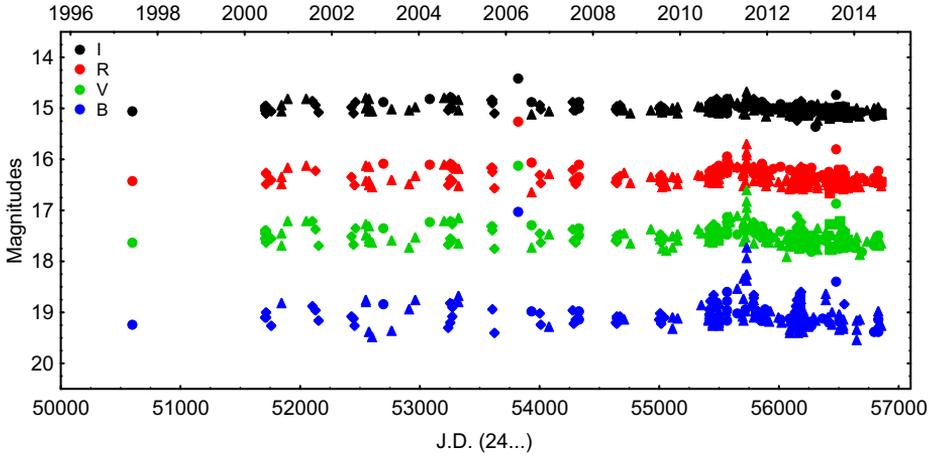, width=0.95\textwidth}}
    \caption[]{$BVRI$ light curves of V1716 Cyg for the period June 1997 -- July 2014}
    \label{countryshape}
  \end{center}
\end{figure}

\begin{table}[htb!!!]
  \begin{center}
  \caption{Photometric CCD observations of V1716 Cyg during the period April 2013 -- July 2014}
  \begin{tabular}{cccccccccccccc}
		  \hline \hline
		  \noalign{\smallskip}
J.D. (24...) & $I$ & $R$ & $V$ & $B$ & Tel & CCD & J.D. (24...) & $I$ & $R$ & $V$ & $B$ & Tel & CCD\\
  \noalign{\smallskip}
  \hline
   \noalign{\smallskip}
56392.510 &	14.97	&	16.38	&	17.44 &	18.64	&	Sch	&	FLI &	56544.264	&	15.08	&	16.42	&	17.69	&	-	&	2-m	&	VA\\
56394.482	&	15.14	&	16.46	&	17.59	&	18.76	&	Sch	&	FLI &	56547.383	&	14.99	&	16.33	&	17.46	&	-	&	Cas	&	FLI\\
56415.418	&	15.09	&	16.52	&	17.61	&	19.12	&	Sch	&	FLI &	56550.363	&	15.03	&	16.37	&	17.58	&	-	&	Cas	&	FLI\\
56428.404	&	15.14	&	16.68	&	17.78	& -	&	Cas &	FLI &			56553.283	&	15.08	&	16.51	&	- &	18.84	&	1.3-m	&	AND\\
56430.408	&	15.00	&	16.33	&	17.28	&	- &	Cas	&	FLI &			56577.306	&	15.19	&	16.58	&	17.58	&	-	& Cas	&	FLI\\
56432.405	&	15.21	&	16.58	&	17.72	&	- &	Cas	&	FLI &			56578.331	&	15.12	&	16.48	&	17.68	&	-	&	Cas	&	FLI\\
56443.367	&	15.00	&	16.41	&	17.50	&	19.02	&	Sch	&	FLI &	56604.264	&	15.07	&	16.45	&	17.73	&	-	&	Cas	&	FLI\\
56444.354	&	15.13	&	16.50	&	17.78	&	19.25	&	Sch	&	FLI &	56636.185	&	15.10	&	16.41	&	17.64	&	-	&	2-m	&	VA\\
56478.363	&	14.74	&	15.81	&	16.87	&	18.41	&	2-m	&	VA &  56655.200	&	15.00	&	16.33	&	17.54	&	19.18	&	Sch	&	FLI\\
56507.273	&	15.13	&	16.49	&	17.76	& -	&	2-m	&	VA &			56656.180	&	15.13	&	16.55	&	17.69	&	19.35	&	Sch	&	FLI\\
56508.318	&	15.03	&	16.29	&	17.54	&	19.19	&	2-m	&	VA &	56657.193	&	15.15	&	16.50	&	17.75	&	19.55	&	Sch	&	FLI\\
56509.288	&	14.99	&	16.35	&	17.46	&	19.10	&	Sch	&	FLI &	56681.189	&	15.16	&	16.56	&	17.89	&	19.16	&	Sch	&	FLI\\
56510.369	&	14.98	&	16.24	&	17.29	&	- &	Cas	&	FLI &			56694.194	&	15.08	&	16.37	&	17.82	&	- &	2-m &	VA\\
56510.386	&	15.05	&	16.41	&	17.54	&	19.06	&	Sch	&	FLI &	56738.590	&	15.04	&	16.43	&	17.62	&	-	&	Sch	&	FLI\\
56511.411	&	15.16	&	16.58	&	17.74	&	19.35	&	Sch	&	FLI &	56799.425	&	15.01	&	16.46	&	17.55	&	- &	Sch	&	FLI\\
56511.413	&	15.16	&	16.55	&	17.60	&	- &	Cas &	FLI &			56801.436	&	15.16	&	16.38	&	17.71	&	19.39	&	2-m	&	VA\\	
56512.398	&	15.09	&	16.44	&	17.60	&	19.32 &	Sch	&	FLI &	56832.391	&	15.06	&	16.23	&	17.50	&	19.16	&	2-m	&	VA\\
56512.403	&	15.05	&	16.40	&	17.64	&	- &	Cas	&	FLI &			56834.356	&	15.09	&	16.39	&	17.67	&	19.34	&	2-m	&	VA\\ 
56513.382	&	15.01	&	16.38	&	17.59	&	- &	Cas	&	FLI &	    56835.462	&	15.15	&	16.42	&	17.71	&	19.39	&	2-m	&	VA\\
56514.350	&	14.99	&	16.41	&	17.62	&	- &	Cas	&	FLI &	    56837.417	&	14.99	&	16.39	&	17.60	&	18.99	&	Sch	&	FLI\\
56517.284	&	15.02	&	16.22	&	17.22	& -	&	Cas	&	FLI &     56838.398	&	15.09	&	16.48	&	17.70	&	19.23	&	Sch	&	FLI\\
56540.274	&	15.13	&	16.59	&	17.75	&	19.28	&	Sch	&	FLI & 56859.390 & 15.12 & 16.45 & - & - & Cas & FLI\\ 
56541.322	&	15.09	&	16.45	&	17.63	&	19.34	&	Sch	&	FLI & 56860.391 & 15.10 & 16.45 & - & - & Cas & FLI\\
56542.381	&	15.01	&	16.39	&	17.61	&	19.15	&	Sch	&	FLI & 56863.339	& 15.09	& 16.53	& 17.67 &	19.27 &	Sch &	FLI\\
56543.396	&	14.97	&	16.21	&	17.41	&	-	&	2-m	&	VA &      56864.356	& 15.12 & - & - & - & Sch	& FLI\\
\hline
  \end{tabular}
  \label{table5}
  \end{center}
\end{table} 

The variations in brightness of V1716 Cyg in the different bands during the period of our photometric study (1997-2014) are 14.43 -- 15.36 mag for $I$-band, 15.28 -- 16.68 mag for $R$-band, 16.14 -- 17.93 mag for $V$-band, and 17.05 -- 19.55 mag for $B$-band. The observed amplitudes are 0.93 mag for $I$-band, 1.40 mag for $R$-band, 1.79 mag for $V$-band and 2.50 mag for $B$-band in the same period.

Figure 5 shows four eruptive events of V1716 Cyg with large amplitudes. The first eruption was detected on March 2006 with an amplitude reaching up to 2.3 mag in $B$-band, the second one was detected on June 2011 with an amplitude of 1.6 mag in $B$-band, the thirdly eruption is detected on September 2012 with an amplitude of 0.7 in $B$-band, and the fourthly eruption is detected on July 2013 with an amplitude about 0.9 mag in $B$-band. These irregular flares can be explain with short-lived accretion-related events at the stellar surface or as flares from UV Ceti type. Other irregular variations of the brightness of V1716 Cyg with smaller amplitudes likely are caused by rotating hot and cool spots on the stellar surface. These results are indication that V1716 Cyg is possibly CTTS with variability of Type I, II and III.

\subsection*{3.5. V2051 Cyg}

V2051 Cyg was discovered as flare star by Parsamian et al. (1994). The star showed burst event on September 7, 1977 with amplitude $\geq$ 4 mag in $U$-band (Parsamian et al. 1994). Poljan\v{c}i\'{c} Beljan et al. (2014) determined the period of the star, which is found to be 384 days.

The $BVRI$ light curves of V2051 Cyg from all our CCD observations (Poljan\v{c}i\'{c} Beljan et al. 2014, and the present paper) are shown in Fig. 6. The symbols used for different telescopes are as in Fig. 2. The photometric results of our CCD $BVRI$ observations of the star are summarized in Table 6. The columns have the same contents as in Table 2. During the period of our observations other flares of the star are not registered, except a few low amplitude increases in brightness in $B$-band. 

\begin{figure}[htb!!!]
  \begin{center}
    \centering{\epsfig{file=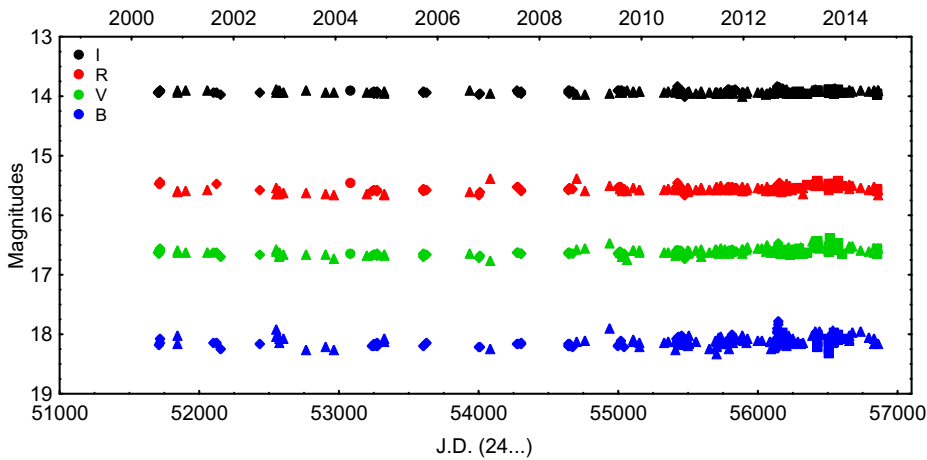, width=0.95\textwidth}}
    \caption[]{$BVRI$ light curves of V2051 Cyg for the period June 2000 -- July 2014}
    \label{countryshape}
  \end{center}
\end{figure}

\begin{table}[htb!!!]
  \begin{center}
  \caption{Photometric CCD observations of V2051 Cyg during the period April 2013 -- July 2014}
  \begin{tabular}{cccccccccccccc}
		  \hline \hline
		  \noalign{\smallskip}
J.D. (24...) & $I$ & $R$ & $V$ & $B$ & Tel & CCD & J.D. (24...) & $I$ & $R$ & $V$ & $B$ & Tel & CCD\\
  \noalign{\smallskip}
  \hline
  \noalign{\smallskip} 
56392.510 &	13.91	&	15.49	&	16.45	& -	&	Sch	&	FLI &			56542.381	&	13.89	&	15.52	&	16.53	&	18.12	&	Sch	&	FLI\\
56394.482	&	13.93	&	15.50	&	16.53	&	18.03	&	Sch	&	FLI &	56547.383	&	13.92	&	15.48	&	16.54	&	17.98	&	Cas	&	FLI\\
56415.418	&	13.89	&	15.51	&	16.55	&	17.96	&	Sch	&	FLI &	56550.363	&	13.96	&	15.52	&	16.58	&	18.13	& Cas	&	FLI\\
56428.404	&	13.93	&	15.49	&	16.55	&	18.02	&	Cas	&	FLI &	56553.283	&	13.88	&	15.49	&	-&	18.07	&	1.3-m &	AND\\
56430.408	&	13.92	&	15.46	&	16.50	&	18.14	&	Cas	&	FLI &	56577.306	&	13.94	&	15.56	&	16.56	&	18.02	&	Cas	&	FLI\\
56432.405	&	13.91	&	15.43	&	16.45	&	18.23	&	Cas	&	FLI &	56578.331	&	13.90	&	15.43	&	16.48	&	18.11	&	Cas &	FLI\\
56443.367	&	13.93	&	15.51	&	16.57	&	18.02	&	Sch	&	FLI &	56604.264	&	13.95	&	15.52	&	16.66	&	18.08	&	Cas	&	FLI\\
56444.354	&	13.91	&	15.46	&	16.49	&	17.97	&	Sch	&	FLI &	56655.200	&	13.91	&	15.51	&	16.54	&	18.12	&	Sch	&	FLI\\
56509.288	&	13.90	&	15.49	&	16.55	&	18.08	&	Sch	&	FLI &	56656.180	&	13.94	&	15.57	&	16.55	&	18.08	&	Sch	&	FLI\\
56510.369	&	13.93	&	15.55	&	16.62	&	18.32	&	Cas	&	FLI &	56657.193	&	13.91	&	15.50	&	16.57	&	18.04	&	Sch	&	FLI\\
56510.386	&	13.91	&	15.52	&	16.54	&	18.07	&	Sch	&	FLI &	56681.189	&	13.93	&	15.51	&	16.54	&	17.99	&	Sch	&	FLI\\
56511.411	&	13.93	&	15.58	&	16.56	&	18.12	&	Sch	&	FLI &	56738.590	&	13.92	&	15.56	&	16.61	&	17.96	&	Sch	&	FLI\\
56511.413	&	13.94	&	15.52	&	16.58	&	18.25	&	Cas	&	FLI &	56799.425	&	13.93	&	15.58	&	16.62	&	18.07	&	Sch	&	FLI\\
56512.398	&	13.91	&	15.51	&	16.55	&	18.06	&	Sch	&	FLI &	56837.417	&	13.92	&	15.54	&	16.57	&	18.08	&	Sch	&	FLI\\
56512.403	&	13.90	&	15.49	&	16.61	&	- &	Cas	&	FLI &			56838.398	&	13.93	&	15.58	&	16.63	&	18.17	&	Sch	&	FLI\\
56513.382	&	13.92	&	15.51	&	16.54	&	18.14	&	Cas	&	FLI &	56859.390 & 13.97 & 15.57 & 16.56 & - & Cas & FLI\\
56514.350	&	13.90	&	15.50	&	- &	18.08	&	Cas	&	FLI &			56860.391 & 13.99 & 15.61 & 16.66 & - & Cas & FLI\\
56517.284	&	13.94	&	15.49	&	16.4	&	- &	Cas	&	FLI &		  56863.339	& 13.91	& 15.58	& 16.57	& 18.17	& Sch	& FLI\\
56540.274	&	13.90	&	15.48	&	16.53	&	17.95	&	Sch	&	FLI & 56864.356	& 13.92	& 15.68 & - & - &	Sch & FLI\\
56541.322	&	13.92	&	15.53	&	16.54	&	18.16	&	Sch	&	FLI & &	&	&	&	&	&	\\
\hline
  \end{tabular}
  \label{table6}
  \end{center}
\end{table} 

The variations in brightness of V2051 Cyg in the different bands during the period of our photometric study (2000-2014) are 13.84 -- 14.02 mag for $I$-band, 15.39 -- 15.68 mag for $R$-band, 16.04 -- 16.77 mag for $V$-band, and 17.80 -- 18.35 mag for $B$-band. The observed amplitudes are 0.18 mag for $I$-band, 0.29 mag for $R$-band, 0.73 mag for $V$-band and 0.55 mag for $B$-band in same period. 

On the basis of the observed flare with amplitude $\geq$ 4 mag in $U$-band by Parsamian et al. (1994), the available literature data and the registered in our study low amplitude variability, we classify V2051 Cyg as low-mass WTTS. Therefore, the variability of the star can be attributed to Type IV.

We are continuing to collect regular $BVRI$ photometric observations of the field of "Gulf of Mexico" in order to classify the stars from our study with higher accuracy. For this purpose of obtaining of spectral observations of the young stellar objects in the field of "Gulf of Mexico" will be of great importance.

\section*{Acknowledgements}

This study was partly supported by ESF and Bulgarian Ministry of Education and Science under the contract BG051PO001-3.3.06-0047. The research has made use of the NASA's Astrophysics Data System Abstract Service.


\end{document}